\documentclass[iop]{emulateapj}

\usepackage[tight]{subfigure}
\usepackage{amsmath}
\usepackage{graphicx}
\usepackage{wasysym}
\usepackage{color}
\usepackage{verbatim}
\usepackage{apjfonts}

\newcommand{\kpc}{{\rm kpc}}

\newcommand{\pc} {{\rm pc}}

\newcommand{\mo}{{\rm M}_\odot}

\newcommand{\gsim}{\lower.7ex\hbox{$\;\stackrel{\textstyle>}{\sim}\;$}}
\newcommand{\lsim}{\lower.7ex\hbox{$\;\stackrel{\textstyle<}{\sim}\;$}}
\newcommand{\mug}{{\rm \mu G}}

\begin{document}

\title{Magnetic fields in cosmological simulations of disk galaxies}
\shorttitle{Magnetic fields in simulations of disk galaxies}
\shortauthors{R.~Pakmor et al.}

\author
{
 R\"udiger~Pakmor\altaffilmark{1}, 
 Federico~Marinacci\altaffilmark{1,2}
 and Volker~Springel\altaffilmark{1,2}
}

\altaffiltext{1}
{
  Heidelberger Institut f\"{u}r Theoretische Studien,
  Schloss-Wolfsbrunnenweg 35, 69118 Heidelberg, Germany
}

\altaffiltext{2}
{
  Zentrum f\"ur Astronomie der Universit\"at Heidelberg,
  Astronomisches Recheninstitut, M\"{o}nchhofstr. 12-14, 69120
  Heidelberg, Germany
}

\begin{abstract}
Observationally, magnetic fields reach equipartition with thermal
energy and cosmic rays in the interstellar medium of disk galaxies
such as the Milky Way. However, thus far cosmological simulations of
the formation and evolution of galaxies have usually neglected
magnetic fields. We employ the moving-mesh code \textsc{Arepo}
to follow for the first time the formation and evolution of a Milky~Way-like 
disk galaxy in its full cosmological context while taking
into account magnetic fields.  We find that a prescribed tiny magnetic seed
field grows exponentially by a small-scale dynamo until it saturates
around $z=4$ with a magnetic energy of about $10\%$ of
the kinetic energy in the center of the galaxy's main progenitor
halo. By $z=2$, a well-defined gaseous disk forms in which the magnetic
field is further amplified by differential rotation, until it
saturates at an average field strength of $\sim 6 \mug$ in the disk
plane.  In this phase, the magnetic field is transformed from a
chaotic small-scale field to an ordered large-scale field
coherent on scales comparable to the disk radius. The final magnetic
field strength, its radial profile and the stellar structure of the
disk compare well with observational data. A minor merger temporarily
increases the magnetic field strength by about a factor of two, before it 
quickly decays back to its saturation value. Our
results are highly insensitive to the initial seed field strength and
suggest that the large-scale magnetic field  in spiral
galaxies can be explained as a result of the cosmic structure
formation process.
\end{abstract}


\section{Introduction}
\label{sec:intro}

Numerous observations have shown that the Universe is highly
magnetized on a variety of scales, ranging from stars to galaxies, and
even up to clusters of galaxies. Among these objects, galaxies are
special because their magnetic energy content appears to be roughly in
equipartition with the thermal energy of the gas and the energy stored
in cosmic rays \citep{Beck1996}. Consequently, magnetic fields are
potentially quite relevant for the dynamical evolution of galaxies, as
well as for the regulation of their star formation rate \citep{Beck2009}.

The origin and evolution of galactic magnetic fields is still not well
understood. Tiny initial magnetic seed fields are assumed to be
generated either in the primordial Universe or by Biermann batteries
in the first stars \citep{Kulsrud2008}. Different processes have been
proposed for subsequently amplifying the magnetic field to the
observed present-day strength, including turbulent dynamo processes
\citep[see, e.g.][]{Kulsrud1997,Arshakian2009,Schleicher2013}, shear flows during
structure formation \citep[see, e.g.][]{Dolag1999}, or a Galactic
dynamo \citep[][for a review see 
\citealt{Widrow2002}]{Hanasz2004}. 
A number of these scenarios have been
studied with numerical magnetohydrodynamics simulations, either
concentrating on isolated disk galaxies
\citep{Hanasz2009,Kotarba2009,Wang2009,Dubois2010,Pakmor2013} or on
larger scales in a cosmological context, ranging from galactic halos
\citep{Beck2012,Beck2013b} to galaxy clusters
\citep{Dolag1999,Dubois2008,Donnert2009} and voids \citep{Beck2013}.

However, until quite recently, the problem of successfully forming
disk galaxies from cosmological initial conditions with properties
similar to observed spiral galaxies has essentially been unsolved,
precluding the possibility of direct cosmological predictions of the
expected magnetic field topology and strength in disk galaxies.  Here,
we present the first high-resolution hydrodynamical simulation of the
formation of a Milky-Way like galaxy in its full cosmological context
including magnetic fields. We make use of the same simulation approach
as in \citet{Marinacci2013} but add magnetic fields that are fully
coupled to the hydrodynamical evolution.  This allows us to produce
realistic disc galaxies while at the same time obtaining predictions
for the magnetic field strength throughout the galaxy as well as its
impact on the global star formation efficiency.  In \S\ref{sec:setup}
we briefly summarize our numerical methodology and specify our initial
conditions.  Results for the amplification of the magnetic field in
the modelled galaxy are discussed in \S\ref{sec:amplification}, while
the present-day structure of the field in the formed disc galaxy is
analyzed in \S\ref{sec:zzero}.  We draw our conclusions in
\S\ref{sec:discussion}.

\begin{figure*}
  \begin{minipage}[b]{0.48\linewidth}
  \centering
  \includegraphics[width=\textwidth]{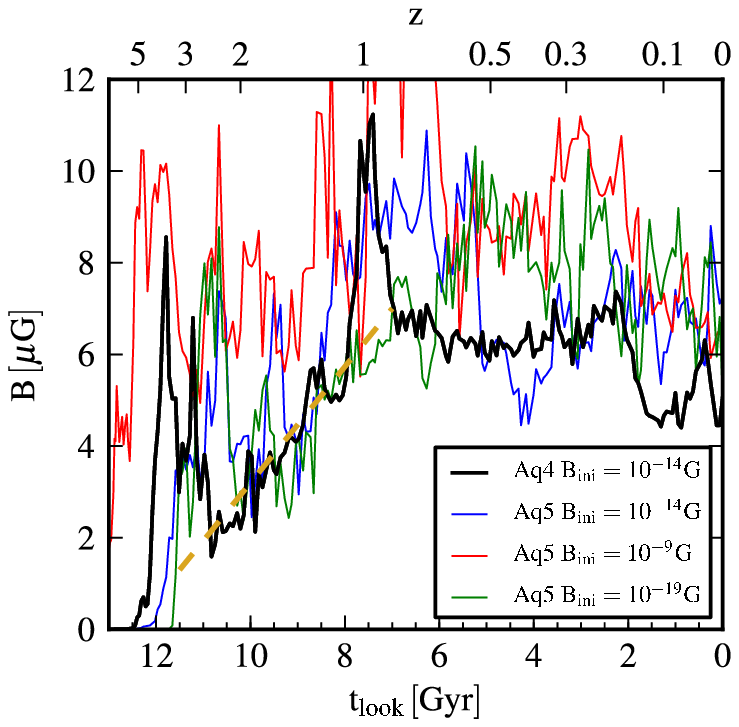}
  \end{minipage}
  \begin{minipage}[b]{0.48\linewidth}
  \centering
  \includegraphics[width=\textwidth]{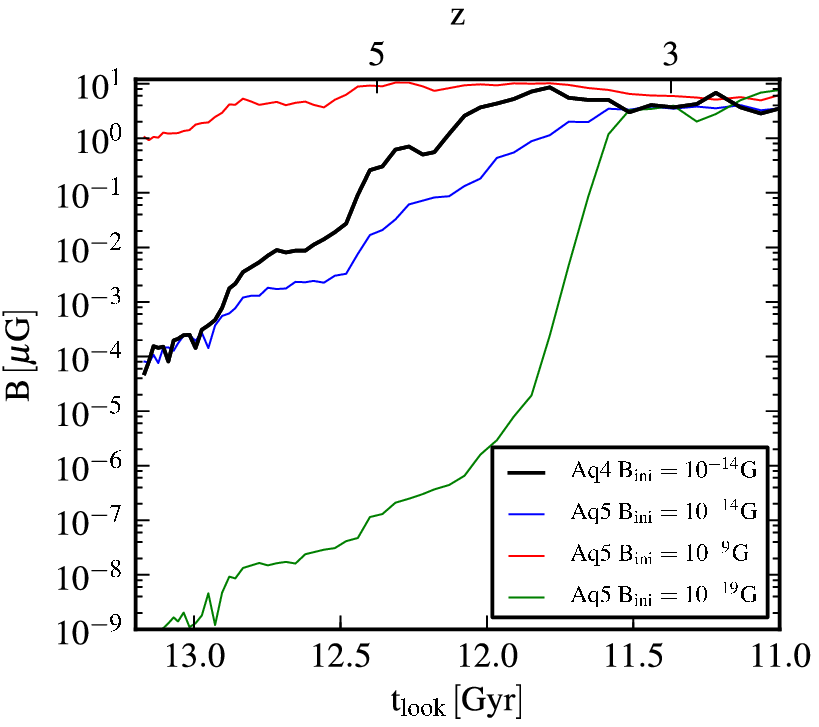}
  \end{minipage}
  \begin{minipage}[b]{0.48\linewidth}
  \centering
  \includegraphics[width=\textwidth]{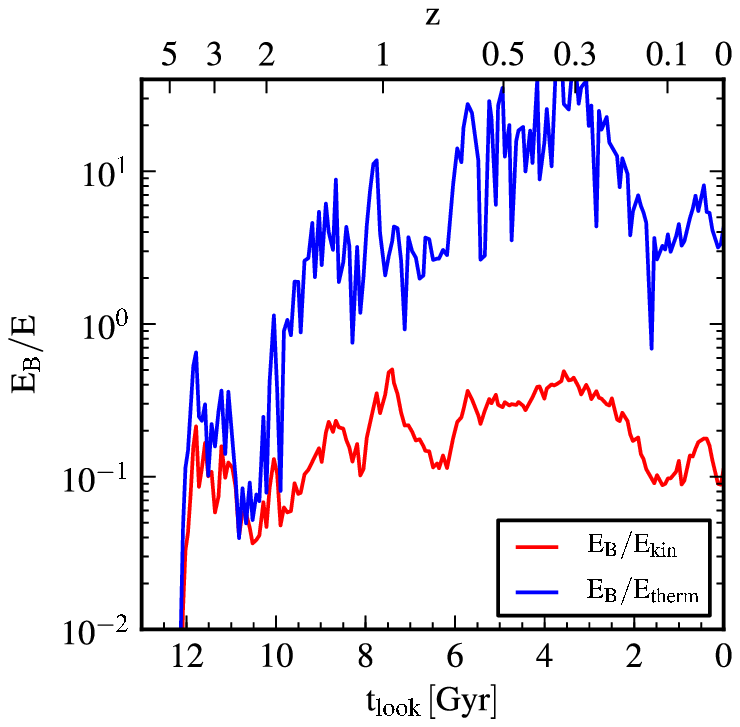}
  \end{minipage}
  \begin{minipage}[b]{0.48\linewidth}
  \centering
  \includegraphics[width=\textwidth]{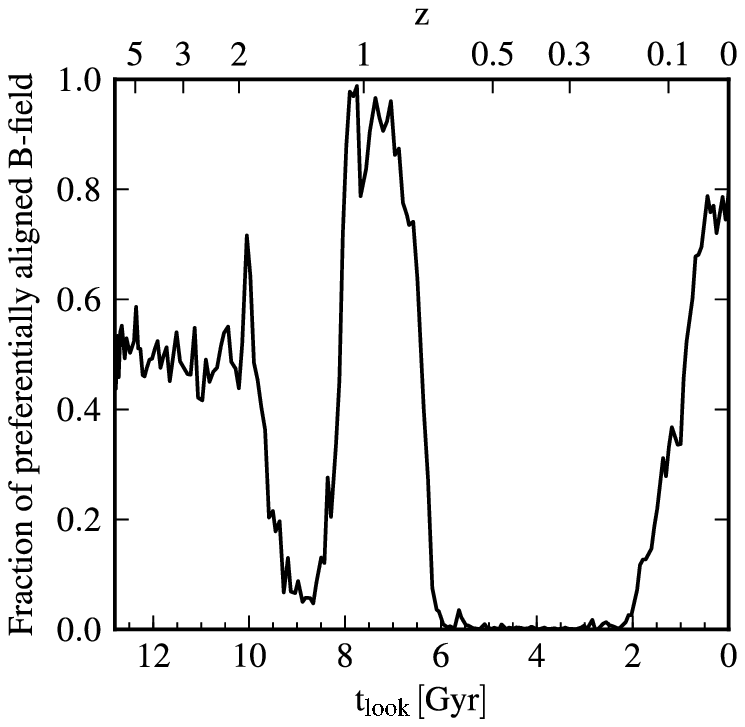}
  \end{minipage}
  \caption{Evolution of the volume weighted average root mean square 
  magnetic field strength (top panels),  the ratio of magnetic energy to kinetic
  energy and thermal energy (bottom left panel), and the volume fraction in which 
  the magnetic field is preferentially aligned with the velocity field (bottom right 
  panel) in a disk with physical radius $15\,\kpc$ and height $2\,\kpc$ centered 
  on the potential minimum of the halo and orientated along the principal axes 
  of all stars within 10\% of the virial radius. The bulk velocity of the galaxy has 
  been removed before calculating the kinetic energy. Although there is no 
  prominent disk at $z > 2$, we use the same volume cut to make the
  comparison easier. The top panels show additional runs with $8$ times
  lower mass resolution (Aq-A-5) and different magnetic seed field strength. The
  yellow line in the top left panel indicates the approximately linear growth of the magnetic field
  strength between $z=2$ and $z=1$.}
  \label{fig:evol}
\end{figure*}

\section{Setup}
\label{sec:setup}

We simulate the formation of a Milky-Way sized galaxy in a
cosmological volume using the ``zoom-in technique''. To this end we
adopt the initial conditions of halo Aq-A-4 from the `Aquarius'
project \citep{Springel2008}. The halo was selected from the `Millenium II'
simulation \citep{MilleniumII} with a mass similar to the Milky Way
and applying a mild isolation criterion to ensure a relatively quiet merger
history, which favors the formation of an extended disk. Indeed, the halo evolves quietly for
most of the time, with the exception of a minor merger around $z=1$.
Our simulation features a baryonic mass
resolution of $5 \times 10^4~\mo$, a dark matter mass resolution of
$3.2 \times 10^5~\mo$ and a gravitational softening of $340~\pc$.  At
this resolution, our simulation is comparable to some of the best
resolved cosmological galaxy formation simulations available today.

We employ the moving-mesh code \textsc{Arepo} \citep{Arepo} and its
new ideal MHD implementation
\citep{Pakmor2011,Pakmor2013}. \textsc{Arepo} combines advantages of
Lagrangian codes (e.g.~small advection errors and automatic
adaptivity) with advantages of Eulerian codes (e.g.~accurate gradient
estimates, fast convergence rate, and no need for an artificial
viscosity), making it a particularly well-matched technique for the
problem at hand.  Our model for baryonic physics \citep[see][for a full
  description]{Vogelsberger2013} includes primordial
and metal line cooling, a sub-grid model for the interstellar medium
(ISM) and star formation \citep{Springel2003}, a self-consistent treatment of stellar
evolution and chemical enrichment, as well as galactic-scale winds
\citep[implemented with a kinetic scheme similar to][]{Puchwein2013} and
supermassive black hole growth and feedback.  The code's configuration and the
setup of the employed baryonic physics modules is identical to the
suite of simulations presented in \citet{Marinacci2013}, to which we refer
for further details, except for the addition of magnetic fields.

We introduce the magnetic field in the initial conditions of the simulation
in terms of a homogeneous comoving\footnote{$B_{\rm{comoving}} = a^{2}\, B_{\rm{physical}}$, where $a$ is the 
scale factor.} seed field with a strength of $10^{-14}\,\rm{G}$, and follow its
evolution in the ideal MHD approximation. At the initial redshift
$z=127$ this is equivalent to a physical magnetic field strength of
$2\times 10^{-5}\mug$.  We note that when stars form from a gas cell as part of the
subgrid model, we assume that the corresponding magnetic field flux is
locked up into the star and removed with the gas. Compared to not
removing the magnetic field \citep[as done in][]{Pakmor2013} this
assumption leads to slightly smaller (by at most a few tens of
percent) magnetic field strength. Also note that this can introduce a
small divergence error into the local magnetic field, however, our MHD
implementation is robust enough to quickly eliminate any such local
error.

\section{Amplification of the magnetic field}
\label{sec:amplification}

The time evolution of the average magnetic field strength in the disk
is shown in Fig.~\ref{fig:evol} (top left panel). The field strength grows
exponentially until it reaches a few $\mug$ around $z=4$. At this
time, the magnetic energy attains about $10\%$ of the kinetic, and
$20\%$ of the thermal energy. Such a large magnetic field already at
high redshift is consistent with recent observations
\citep{Kronberg2008}. Even though strong fluctuations occur, we
note there is no significant further net amplification of the magnetic
field and no evident large-scale ordering before the galaxy forms a
well-defined disk at about $z=2$.

\begin{figure*}
  \centering
  \includegraphics[width=0.82\linewidth]{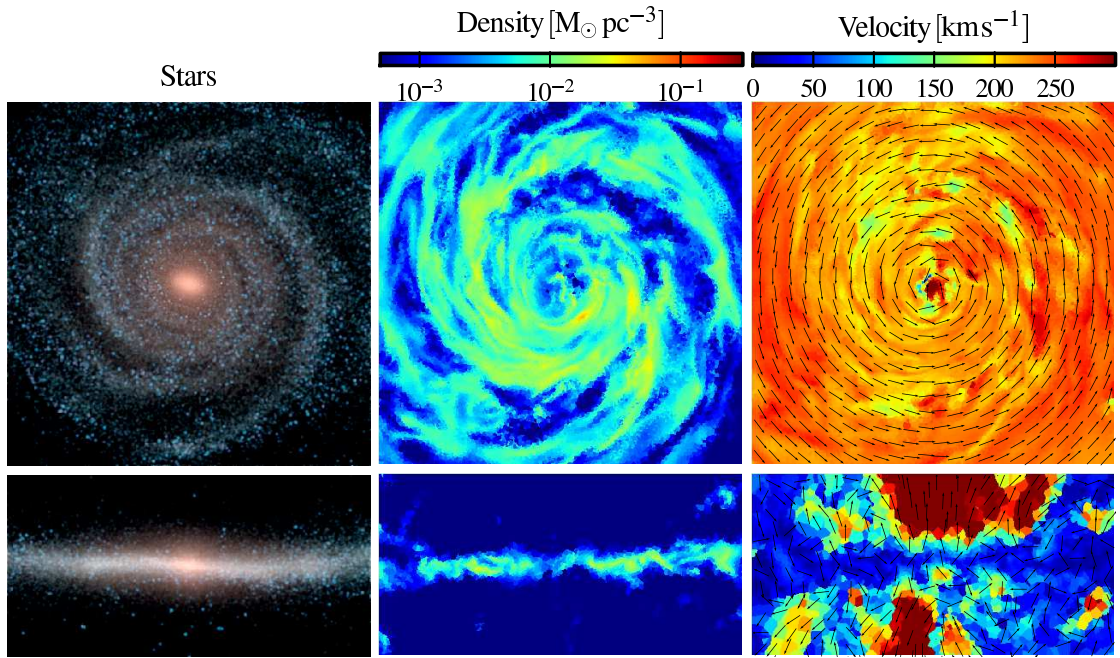}
  \caption{The stellar and gaseous disk at redshift $z=0$. The column
    on the left shows the projected stellar density using a
    logarithmic mapping of $K$-, $B$- and $U$-band luminosity of the
    stars to a RGB-composite. The middle and right columns display,
    respectively, slices through the center of the galaxy of the gas
    density and the horizontal (top panel) and vertical (bottom panel)
    velocity fields.  The coordinate
    system shown has been oriented along the principal axes of all
    stars within 10\% of the virial radius, and the bulk velocity of
    the galaxy has been removed. Top and bottom rows give face-on and
    edge-on views, respectively. The panels on the right include
    overlaid arrows to indicate the velocity vector field. The region
    shown is a box with a width of $50~\kpc$.}
\label{fig:disk}
\end{figure*}

\begin{figure*}
  \centering
  \includegraphics[width=0.82\linewidth]{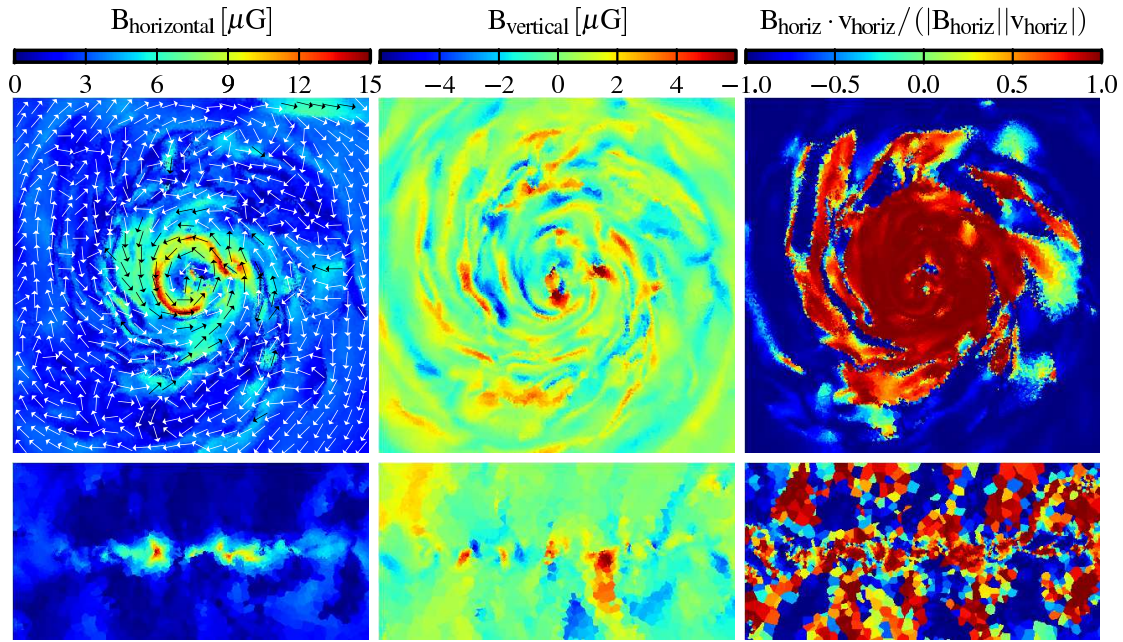}
  \caption{The magnetic field in the disk at redshift $z=0$. Similar
    to Fig.~\ref{fig:disk}, but now showing slices of the magnetic
    field strength in the disk plane (left column), the component of
    the magnetic field perpendicular to the disk plane (middle
    column), and the orientation of the magnetic field relative to the
    orientation of the velocity field (right column). In the top left
    panel, black and white arrows have been overlaid to
    indicate the direction of the magnetic field in the disk.}
\label{fig:diskB}
\end{figure*}

This early evolution of the magnetic field is consistent with a
turbulent small-scale dynamo seen in previous simulations of magnetic
field growth in cosmological halos \citep{Dolag1999,Beck2012} and
predicted analytically \citep{Schober2013}.  However, in contrast to
the analytical models the main driver of turbulence in the ISM of our
simulation are accretion flows onto the forming galaxy. We do not take
supernovae into account as a source of turbulence, because our
sub-scale model for star formation and supernova feedback does not
treat them explicitly \citep{Springel2003}.  It is thus particularly
interesting that our primordial magnetic seed field is amplified to
saturation already at high redshifts through structure growth
processes alone, indicating that supernova feedback may not be
necessary to amplify fields to $\mug$ strength in young galaxies. This
may be important in particular for the magnetic field amplification in
dwarf galaxies. We stress that the strength at which the
magnetic field saturates in this early evolution does not depend on
the size or orientation of the initial seed field. As Fig.~\ref{fig:evol} 
(top right panel) shows, saturation
is always reached before $z=3$ and the saturation strength depends 
only very weakly on the initial seed field because of the fast exponential 
amplification, in agreement with the findings in \citet{Beck2012}.
Only a too large initial magnetic field will cause a problem, since
it can lead to a larger saturation strength.

After the disk has formed around $z=2$, the magnetic field strength
grows approximately linearly with time, until it again saturates, at
an average field strength of about $6\mug$. This linear growth due to
differential rotation in the disk is the same as found in simulations
of isolated disk galaxies \citep{Hanasz2009,Dubois2010,Pakmor2013}.
For a simulation with identical initial conditions but eight times
lower mass resolution (run Aq-A-5 in the Aquarius naming convention),
we find that the final field strength is essentially the same,
suggesting that this quantity is quite well converged (Fig.~\ref{fig:evol}
top right panel). However, the magnetic field strength fluctuates significantly more
in the lower resolution runs.

Shortly after $z=2$, the magnetic energy has reached equipartition
with the thermal energy as a result of an increase of the magnetic
energy as well as a decrease of the thermal energy. It also reaches
about $20\%$ of the kinetic energy at $z=1$. For the following
$8\,\rm{Gyrs}$, $E_{\rm B} / E_{\rm kin}$ and $E_{\rm B} / E_{\rm
  thermal}$ fluctuate between $0.2$ to $0.5$ and $3$ to $20$,
respectively (see Fig.~\ref{fig:evol}
bottom left panel).

The differential rotation in the gas disk also orders the magnetic
field on the scale of the disk. Once the disk has formed, the magnetic
field is transformed on a timescale of about $2\,\rm{Gyrs}$ from a
randomly oriented small-scale field to a configuration in which it is
ordered on scales comparable to the disk, with an alignment along the
azimuthal velocity field. Since the ideal MHD equations are symmetric
under a global sign flip of the $B$-field, we do not expect that
either the parallel or anti-parallel orientation relative to the
velocity field is preferred. Nevertheless, one of these directions
typically ends up prevailing over the other, with most of the disk
showing either the parallel or anti-parallel orientation.  Moreover,
the dominant orientation is not necessarily permanent, but can reverse
on a typical timescale of about $2\,\rm{Gyrs}$ if the disk is
sufficiently perturbed. We show this in Fig.~\ref{fig:evol}
  (bottom right panel), where we plot the fraction of the magnetic
  field aligned with the gas velocity field.  There is no dominant
  alignment until about $z=2$, but then the field becomes almost
  completely anti-aligned with the velocity field at $z\sim 1.5$,
  followed by a reversal of the orientation and an eventual switch
  back to the anti-aligned orientation on a timescale of about
  $2\,\rm{Gyrs}$ between each reversal. After this phase, the
  $B$-field stays completely anti-aligned for more then
  $4\,\rm{Gyrs}$, but starts changing again in the last $2\,\rm{Gyrs}$
  to reach a predominantly aligned orientation at $z=0$.

We note that the average magnetic field strength peaks close to $z=1$
when a minor merger occurs \citep[see][]{Marinacci2013}. In this
merger, the mean field strength is amplified by almost a factor of
two, but then decays back to its prior saturation value in about one
Gyr. An amplification during mergers has also been observed in
dedicated simulations of isolated galaxy mergers 
\citep{Kotarba2011,Geng2012}.

\section{The galaxy at z=0}
\label{sec:zzero}

\begin{figure*}
  \begin{minipage}[b]{0.45\linewidth}
  \centering
  \includegraphics[width=\textwidth]{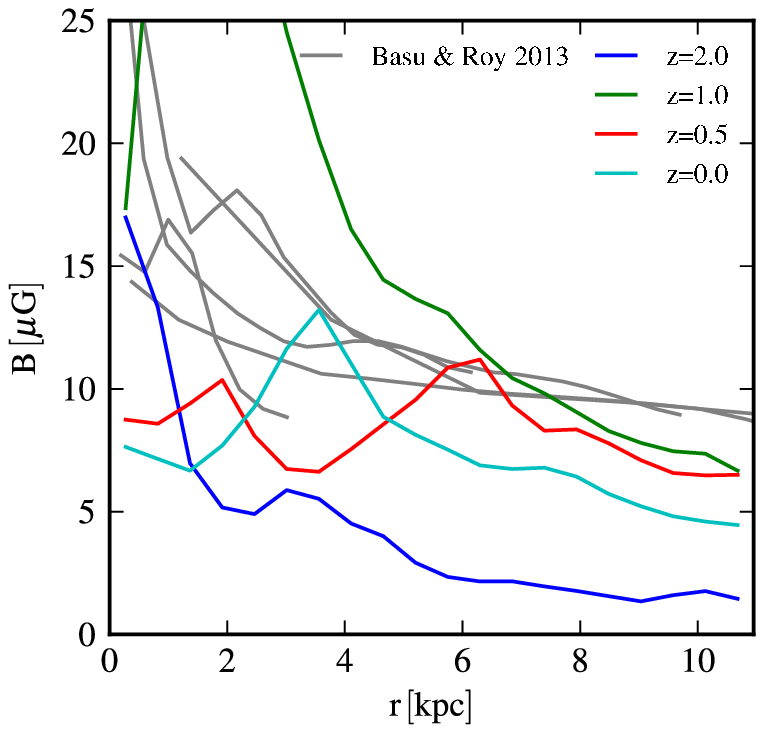}
  \end{minipage}
  \hspace{0.5cm}
  \begin{minipage}[b]{0.45\linewidth}
  \centering
  \includegraphics[width=\textwidth]{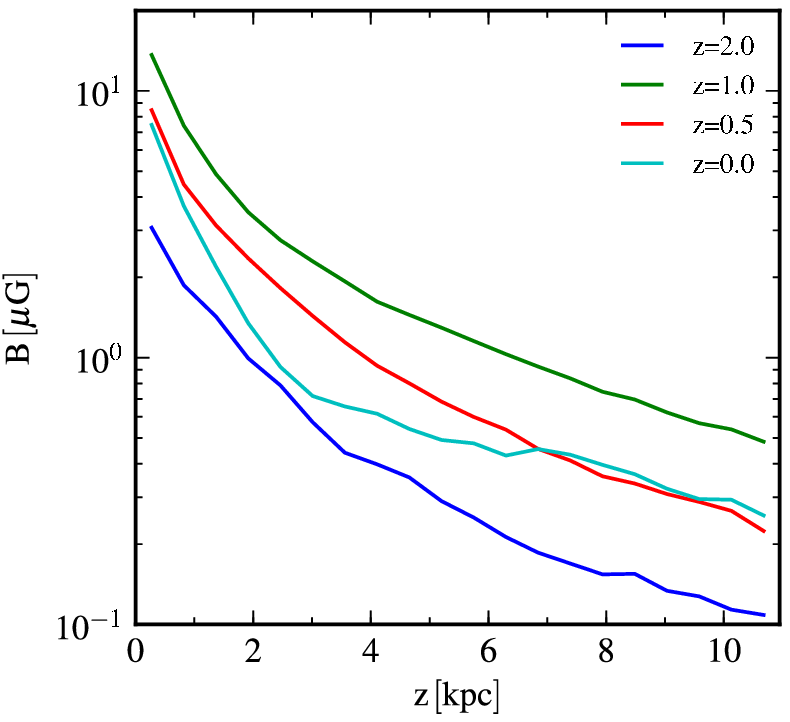}
  \end{minipage}
  \caption{Magnetic field strength profiles in the radial (left panel)
    and $z$ directions (right panel).  The panels show the
    volume-weighted average root mean square $B$-field strength, using
    radial bins with a total height of $1~\kpc$ centered on the disk
    plane for the radial profile, and using height bins at a constant
    radius of $15~\kpc$ for the vertical profile averaged over both
    sides of the disk. The grey lines in the left panel show the field
    strength profiles inferred by \citet{Basu2013} for five observed
    galaxies. }
\label{fig:bfield}
\end{figure*}

The projected stellar density as well as slices through the gas
density and velocity fields in the disk plane at $z=0$ are shown in
Fig.~\ref{fig:disk}. Similarly to the non-MHD runs discussed in detail
in \citet{Marinacci2013}, a rotationally supported disk galaxy with
realistic stellar mass and scale length has formed. The gas shows a
very regular circular velocity field in the disk plane, and spiral
structures are visible in the density and velocity fields. The
vertical velocity and density structures are dominated by irregular
bipolar outflows as a result of our kinetic wind implementation, 
which launches winds preferentially along the galaxy's spin axis.

Fig.~\ref{fig:diskB} presents the magnetic field configuration of our
galaxy at $z=0$. The azimuthal magnetic field in the disk is mostly
oriented parallel to the velocity field (see also Fig.~\ref{fig:evol}
bottom right panel). Its sign relative to the
direction of the velocity field changes in magnetic spiral arms that
correlate with the spiral structure of the density field.  The
orientation of the magnetic field perpendicular to the disk exhibits
no clear correlation with the velocity field and changes on scales of
order the vertical disk scale height.

The magnetic field strength reaches about $15\,\mug$ close to the
center of the disk and a few $\mug$ in the outer parts, in good
agreement with the magnetic field inferred for observed spiral
galaxies \citep[see, e.g.][]{Beck1996,Beck2009,Basu2013}. The magnetic
field component perpendicular to the disk is significantly smaller
compared with the azimuthal component and also changes sign in spiral
arms. The magnetic field strength in the very center of the disk is
comparable to values at larger radii of the order of one $\kpc$, because the
central supermassive black hole included in our simulations blows out
dense, highly magnetized gas from the galactic centre, lowering the
gas density and magnetization there. This may potentially cause an
underestimate of the innermost field strength if this outflow is overly
strong.

The radial and vertical profiles of the magnetic field strength for
different redshifts between $z=2$, when the disk first forms, and
$z=0$ are shown in Fig.~\ref{fig:bfield}. At $z=2$, the field reaches
about $15~\mug$ close to the center of the disk and then declines
roughly linearly with radius to a value of only $1~\mug$ at radii
larger than a few kpc. At $z=1$, the radial magnetic field has grown
at all radii. It reaches $30~\mug$ close to the center, and $7~\mug$ to
$15~\mug$ at radii between $5~\kpc$ and $10~\kpc$ due to a minor
merger at that redshift.  At $z=0.5$, the magnetic field has dropped
again at radii smaller than $6~\kpc$, and most notably peaks now
around a radius of $6~\kpc$ with a field strength of only about
$10~\mug$, as the temporary amplification triggered by the minor
merger fades away.  From $z=0.5$ to $z=0$ the magnetic field hardly
evolves. As shown in Fig.~\ref{fig:bfield}, the radial profile of the
field strength at $z=0$ agrees well with profiles observed for some
spiral galaxies \citep{Basu2013}.  In the vertical direction, the
magnetic field strength declines exponentially, quickly dropping below
$1~\mug$ at heights larger than a few kpc.  The overall time evolution
of the vertical profile follows closely that found for the radial
profile.

\section{Conclusion}
\label{sec:discussion}

The hydrodynamical cosmological simulation presented here is the first
successful formation of a present-day Milky Way-like disk galaxy in
which the dynamics of magnetic fields has been included.  Starting
with a tiny seed field, we find that the field is amplified first by
an exponential small-scale dynamo, and later by differential rotation
in the disk after it formed. The field strength grows roughly to
equipartition in the ISM but has only a minor effect on the global
star formation history of the galaxy, at least at late times. Compared
with the corresponding non-MHD simulations of \citet{Marinacci2013},
we find an equally well defined disk morphology, and almost the same
stellar mass.

Interestingly, the strength and shape of the magnetic field predicted
by our simulations agrees rather well with observational data.  Given
the lack of any free parameters in our modelling of the magnetic field
(recall that the results are invariant with respect to the seed field
strength over a very wide range), this is a non-trivial outcome that
shows that large-scale galactic fields can in principle be well
understood as a result of structure growth alone.  We also note that
our results are numerically robust and converged, although we concede
that the unresolved small-scale dynamics in the ISM remains a source
of uncertainty.

The full three-dimensional magnetic field structure we predict in our
simulation can in principle be used to refine models for the
propagation of charged particles in galaxies such as the Milky Way. In
fact, it is a tantalizing prospect to compute yet much higher
resolution models in the future, and then to carry out detailed
comparisons between the simulated magnetic field properties both with
observations and models of the magnetic field of the Milky Way
\citep{Jansson2012a,Jansson2012b,Oppermann2012}. This
will allow us to significantly improve our understanding of the
evolution of galactic magnetic fields in the Universe.  Moreover,
the ability to simulate magnetic fields in galaxy formation allows
the addition of cosmic rays and of a proper treatment of their transport along
the field lines. Such a consistent treatment will be needed to accurately
study the mechanism and relevance of cosmic ray driven winds
\citep[see, e.g.][]{Ptuskin1997,Uhlig2012,Hanasz2013}, which 
have been conjectured to be a highly important feedback mechanism 
especially in small galaxies.

\section*{Acknowledgements}
We thank the referee for a constructive and critical report.
F.M.~acknowledges support by the DFG Research Centre SFB-881 `The
Milky Way System' through project A1. R.P.~acknowledges support by the
European Research Council under ERC-StG grant EXAGAL-308037. V.S
thanks for support by Priority Programme 1648 `SPPEXA' of the German
Science Foundation, and by the Klaus Tschira Foundation. Simulations
of this paper used the SuperMUC computer at the Leibniz Computing
Centre, Garching, under project PR85JE of the Gauss Centre for
Supercomputing, Germany.

\bibliographystyle{apj}
%\bibliography{letter}

\end{document}